\documentclass[twocolumn,showpacs,pra,amsmath,amssymb,floatfix]{revtex4}
\usepackage{amsmath,amssymb}
\usepackage{bm}
\usepackage{epsfig}
\usepackage{psfrag}

\newcommand{\bd}{\bm}

\begin{document}

\title{Landau functions for non-interacting bosons}


 \author{Andreas Sinner, Florian Sch\"{u}tz, and Peter Kopietz}
  
  \affiliation{Institute f\"{u}r Theoretische Physik, Universit\"{a}t
    Frankfurt,  Max-von-Laue Strasse 1, 60438 Frankfurt, Germany}

\date{May 17, 2006}

 \begin{abstract}

We discuss the statistics 
of Bose-Einstein condensation (BEC)
in a canonical ensemble of $N$ non-interacting bosons
in terms of a Landau function ${\cal{L}}^{\rm BEC}_N ( q )$
defined by
the logarithm of the probability distribution of the
order parameter $q$ for BEC.
We also discuss the corresponding Landau function
for  spontaneous symmetry breaking (SSB), which
for finite $N$ should be distinguished from 
${\cal{L}}_N^{\rm BEC} ( q )$.
Only for  $N \rightarrow \infty$  BEC and SSB can be described by the same 
Landau function which depends on the
dimensionality and on the form of  the external
potential in a surprisingly complex manner.
For bosons confined by a three-dimensional
harmonic trap the Landau function exhibits the 
usual behavior expected for continuous phase transitions.

\end{abstract}



  \maketitle
\begin{figure}[tb]    
   \centering
 \psfrag{t1}{$T >T_c$}
 \psfrag{t2}{$T = T_c$}
 \psfrag{t3}{$T < T_c$}
 \psfrag{p}{$ p $}
 \psfrag{L}{${\cal{L}}$}
      \epsfig{file=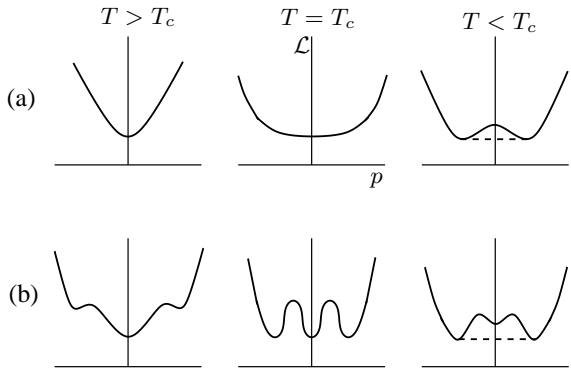,width=75mm}
  \caption{%
Typical evolution of the Landau function ${\cal{L}}$ as a function of the
relevant order parameter $p$
close to  (a) second order and  (b) first order  phase transitions.
The dashed line denotes the plateau  of the 
corresponding Gibbs potential for $T < T_c$.
}
    \label{fig:Landau}
  \end{figure}

The  concept of a Landau function ${\cal{L}} ( p )$
has been extremely useful to understand the nature of phase transitions \cite{Goldenfeld92}.
For example, in the vicinity of a second order (continuous)
thermal phase transition
${\cal{L}} ( p )$ 
smoothly develops minima at non-zero values
of the relevant order parameter $ p$ when the temperature $T$ drops below
the critical temperature $T_c$, see Fig. \ref{fig:Landau} (a).
On the other hand,
in the case of a first order phase transition
the location of the
minima of ${\cal{L}} (  p  )$
changes discontinuously at $T = T_c$, as shown in  Fig. \ref{fig:Landau}(b).
While for infinite systems and in the absence 
of spontaneous symmetry breaking (SSB)
${\cal{L}} ( p )$
can be identified with the
Gibbs potential, this
identification is in general not valid
for finite systems or in the symmetry broken phase.
In the latter case
the Gibbs potential, which by construction
is a convex function of $p$,  exhibits a plateau
as shown in  Fig.~\ref{fig:Landau}.
Microscopically, 
${\cal{L}} ( p )$  is defined in terms of
the logarithm of the
probability distribution of the order parameter
in a given statistical ensemble~\cite{Goldenfeld92}.
From this
probabilistic definition 
it is clear that ${\cal{L}} ( p )$  does not have
a strictly thermodynamic interpretation.

For non-interacting systems the Landau function is
in most cases not very interesting, because the correlations responsible
for phase transitions
are usually due to interactions between the microscopic
degrees of freedom. 
An important exception are non-interacting bosons,
where at low temperatures
the correlations imposed by quantum statistics 
are sufficient to give rise
to a phase transition, the  
Bose-Einstein condensation (BEC) \cite{Pitaevskii03,Leggett01}.
In the condensed state the  fraction of particles
that occupy the single-particle state with the lowest energy
is of the order of unity. More precisely,
the  order parameter for BEC can be chosen to be
the expectation value of the operator
$
 \hat{q} =   b^{\dagger}_0 b_0  /{ N }
$,
where $b_0$ annihilates a boson in the single-particle ground state and
$N$ is the total number of particles in the system.

Conceptually, BEC should be distinguished from SSB,
which manifests itself via a finite expectation value of the 
operator
 $
 \hat{\phi} =  b_0  /\sqrt{{N}}
 $ in the limit $ N \rightarrow \infty$.
Of course, for any finite $N$ there is no off-diagonal long-range order,
so that  $ \langle \hat{\phi} \rangle =0$. 
For interacting bosons
in the limit $N \rightarrow \infty$ it can be shown
that $\langle \hat{q} 
 \rangle  =  | \langle \hat{\phi  } 
 \rangle | ^2$ 
whenever the Bogoliubov approximation of replacing
the operator $b_0 $ by a complex number 
yields the right pressure~\cite{Suto05}. 
For finite systems, however, one should distinguish between
SSB and BEC.  More precisely, both phenomena
should be characterized by different probability distributions,
$P_N^{\rm SSB} (\phi )$ and $P_N^{\rm BEC} ( q )$,
which depend on the eigenvalues $q$ and $\phi$ of the
operators $ \hat{q}$ and $\hat{\phi}$ defined above.
We parameterize the probability distributions in terms of
Landau functions ${\cal{L}}^{\rm SSB}_N ( \phi )$ and
${\cal{L}}^{\rm BEC}_N ( q )$,
see Eqs.~(\ref{eq:PNSSB},\ref{eq:PNBEC}) below.
For {\it{interacting}} bosons the condensation transition
is known to be second order~\cite{Pitaevskii03}, so that the
Landau functions resemble in this case   
Fig.~\ref{fig:Landau} (a).
But how do the Landau functions of {\it{non-interacting}} bosons look like?
Because the order parameter is continuous,
the first order scenario in Fig.~\ref{fig:Landau} (b) can be ruled out.
In spite of the
enormous activity in the field of  BEC in recent years,
this question has apparently not been addressed in the literature.
Although the probability distribution
$P_N^{\rm BEC} ( q )$ has been studied by several 
authors \cite{Grossmann97,Weiss97,Balazs98,Holthaus99,Kocharovsky00,Wilkens00},
the associated Landau function ${\cal{L}}^{\rm BEC}_N ( q )$ has not been 
considered. The corresponding
quantities for SSB  have not even been 
properly defined in the literature.

To describe  BEC experiments in finite clusters
of bosonic atoms~\cite{Dalfovo99}, we
use a statistical ensemble with constant particle number $N$, i.e., the
canonical ensemble.
The mathematically more convenient grand canonical ensemble is neither
relevant for experiments \cite{Dalfovo99}, nor
is it appropriate to describe the  condensed phase~\cite{Ziff77,Balazs98}.
To derive explicit expressions for 
${\cal{L}}^{\rm SSB}_N ( \phi)$ and
${\cal{L}}^{\rm BEC}_N ( q )$,
consider the canonical partition function,
 $
 Z_N = {\rm Tr}_N  \exp [ - \beta \hat{H}  ]
 $,
where  $\beta =1/T$ is the inverse temperature and
${\rm Tr}_N$ denotes the trace over the $N$-particle Hilbert space
of a system of non-interacting 
bosons with Hamiltonian
$ \hat{H} = \sum_{\bd{m}} E_{\bd{m}} b^{\dagger}_{\bd{m}}  b_{ \bd{m}}$.
Here, $b_{\bd{m}}$ annihilates a boson in a single particle state
with quantum numbers $\bd{m}$ and energy $E_{\bd{m}}$.
For free bosons (FB) with mass $M$
confined to a $D$-dimensional volume $L^D $
with periodic boundary conditions
$E_{\bd{m}} = (2 \pi \hbar {\bd{m}} /L )^2 / 2 M $, 
where ${\bd{m}}$ is a $D$-dimensional vector with
integer components $m_i = 0, \pm 1 , \pm 2 , \ldots$, $i =1, \ldots , D$.
For bosons confined by a harmonic  potential (HB)
with  oscillation frequency $\omega$ the energy is
$E_{\bd{m}} = \hbar \omega \sum_{ i =1}^{D} ( m_i + 1/2 )$, where
$m_i = 0 , 1 ,2 , \ldots$.
For later convenience we introduce dimensionless
energies $\epsilon_{\bd{m}} = \beta E_{\bd{m}}$ and 
the relevant dimensionless 
density $\rho = (\lambda_{\rm th} /L)^D N$, where
for FB  the thermal de Broglie wavelength is
 $
\lambda_{\rm th} = \hbar \sqrt{  2 \pi \beta / M}
 $, whereas for HB we define
$\lambda_{\rm th} = \hbar \omega \beta L$ and 
$L = ( M \omega / \hbar )^{1/2}$.

With the help of the projection operator 
 $
 \delta_{ N , \hat{N}} = \int_{0}^{2 \pi} \frac{ d y}{ 2 \pi} e^{ i y ( N - \hat{N} ) }
 $, where
$\hat{N} = \sum_{ \bd{m}} b^{\dagger}_{\bd{m}} b_{\bd{m}}$, we can express
$Z_N$ in terms of a trace over the entire
Fock space~\cite{Politzer96},
\begin{equation}
 Z_N = 
 {\rm Tr}  \left[  \delta_{ N , \hat{N}}  e^{ - \beta \hat{H} } \right] = 
\int_{0}^{2 \pi} \frac{ d y}{ 2 \pi} e^{ i y  N}    {\rm Tr}
 e^{ - \beta \hat{H} - i y \hat{N} }   
 \; .
\end{equation}
The partial trace over the $\bd{m}  \neq 0$ sector of the Fock space
yields
 \begin{eqnarray}
 {\rm Tr}_{ \bd{m} \neq 0}  e^{  -  \sum_{ \bd{m} \neq 0 } ( 
 \epsilon_{\bd{m}} + i y )
 b^{\dagger}_{\bd{m}} b_{\bd{m}} }  
  =
e^{  - \gamma^{-1}   J_D ( iy , \gamma )  }
 \; ,
 \end{eqnarray}
where $\gamma = \rho /N$ and
 $J_D (\alpha  , \gamma ) =   \gamma  \sum_{\bd{m} \neq 0 }
 \ln [  1 -  e^{- \epsilon_{\bd{m}}  - \alpha } ]$.
The contribution from the $\bd{m}  = 0$ sector  
can be written in terms of
coherent states $| z  \rangle$ satisfying $b_0 | z \rangle = z | z \rangle$  as follows,
 \begin{eqnarray}
 {\rm Tr}_{ \bd{m} = 0} e^{  -  ( \epsilon_0 + i y) 
 b^{\dagger}_{0} b_{0} }  
 & = & \int \frac{ d^2 z}{\pi } e^{ - | z |^2 }  \langle z |  
e^{ -  ( \epsilon_0 + i y ) 
 b^{\dagger}_{0} b_{0} }    | z  \rangle
\nonumber
 \\
 &   & \hspace{-15mm} = \int \frac{ d^2 z}{\pi } \sum_{ n =0}^{\infty} \frac{ | z |^{2n}}{ n !}
 e^{ - | z |^2 } e^{ - ( \epsilon_0 + i y) n }
 \; ,
 \end{eqnarray}
where $d^2 z = d {\rm Re} z  d {\rm Im} z$, and
in the second line we have evaluated the matrix element
by  inserting the resolution of unity in the basis $ | n \rangle$ of particle number eigenstates,
$ b^{\dagger}_0 b_0 | n \rangle = n | n \rangle$.
With $ \phi = z / \sqrt{N}$ we finally obtain
 \begin{equation}
 Z_N = \int d^2 \phi e^{ - N  {\cal{L}}^{ \rm{SSB} }_N ( \phi ) }
 \; ,
 \end{equation}
where the Landau function for SSB is
 \begin{equation}
{\cal{L}}^{\rm{SSB}}_N ( \phi  ) 
=  | \phi |^2 - \frac{1}{N} \ln \Bigl[
 \frac{N}{\pi}
 \sum_{ n =0}^{N} 
 \frac{ (N  | \phi |^2 )^n }{n !} e^{ - N {\cal{L}}_N^{\rm{BEC}} ( n/N ) }
 \Bigr]
 ,
 \label{eq:Lphires}
 \end{equation}
and the corresponding Landau function for BEC is
 \begin{equation}
 {\cal{L}}_N^{\rm BEC} ( q ) = \epsilon_0 q - \frac{1}{N} \ln \Bigl[
\int_{ 0 }^{2 \pi} \frac{ dy}{2 \pi}  e^{    i y  N ( 1- q )  -  \frac{N}{\rho}  J_D ( iy , \frac{\rho}{N} )  }
 \Bigr]
 \; .
 \label{eq:Lnres}
 \end{equation}
We have normalized the Landau functions such
that they are dimensionless and have finite limits for $N \rightarrow \infty$.
We conclude that the normalized  
probability density of the order parameter $\phi $ 
for SSB  is    
 \begin{equation}
 P_N^{\rm SSB} ( \phi ) = Z_N^{-1}  e^{- N {\cal{L}}_N^{\rm SSB} ( \phi ) }
 \; ,
 \label{eq:PNSSB}
 \end{equation}
while the probability of observing a fraction  $q = n /N$ 
of bosons in the single-particle state with lowest energy is
 \begin{eqnarray}
 P_N^{\rm BEC} ( q ) 
 & = &
Z_N^{-1} e^{ - N {\cal{L}}_N^{\rm BEC} ( q ) }
 \; .
 \label{eq:PNBEC}
 \end{eqnarray}
For HB the probability 
$P_N^{\rm BEC} (q)$ has been calculated numerically
for $N \leq 500$ 
by Balazs and Bergeman~\cite{Balazs98} using an efficient 
recursive algorithm; 
however, they  neither considered
the corresponding Landau function
nor the analogous quantities for SSB.
 
\begin{figure}[tb]    
   \centering
 \epsfig{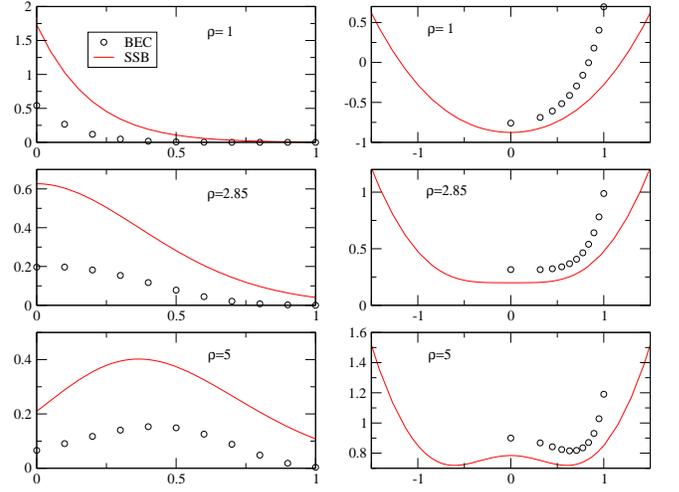}
  \caption{%
(Color online) Order parameter probability distributions (left column) and 
Landau functions
(right column)
of $N=10$ bosons in a three-dimensional harmonic potential for
different dimensionless densities 
$\rho = ( \lambda_{\rm th} / L )^3 N$.
The plots in the second row correspond to
the critical density  $\rho_c \approx 2.85$ for $N=10$.
Discrete points denote ${P}_N^{\rm BEC}$ (plotted versus $q=n/N$) and
${\cal{L}}_N^{\rm BEC}$ (versus  $\sqrt{q}$), 
while the continuous curves
represent $P_N^{\rm SSB}$ (versus $ | \phi |^2$) and
${\cal{L}}_N^{\rm SSB}$ (versus $ {\rm Re} \phi $).
}
    \label{fig:Pfinite}
  \end{figure}

In Fig.~\ref{fig:Pfinite} we show numerical results 
for  the above Landau functions  and the
associated probability distributions for HB in $D=3$.
Obviously, the behavior is typical for  second order phase transitions.
We shall show shortly that for HB in $D=3$ this remains
true for $N \rightarrow \infty$.
In this limit the critical (dimensionless) density for BEC and SSB
in $D$ dimensions is $\rho_c =  \zeta ( d ) $, 
where $d = D/2$ for FB and $d=D$ for HB.
For finite clusters with $N \geq 2$ particles
we define the critical density $\rho_c^{\rm BEC}(N)$
for BEC as the value of $\rho$ where
${\cal{L}}_N^{\rm BEC} ( 0 ) = {\cal{L}}_N^{\rm BEC} ( 1/N)$.
Analogously, 
 at $ \rho= \rho_c^{\rm SSB}(N)$ 
the minimum of  
${\cal{L}}_N^{\rm SSB} ( \phi )$ shifts from $\phi=0$ to
$ | \phi | > 0$. This  definition of 
the critical density (and the associated critical temperature)
in finite clusters is free of ambiguities
and remains valid for interacting 
bosons \cite{Grossmann97,Weiss97,Kocharovsky00,Wilkens00}.
Of course, in finite systems one may use other criteria, such as the behavior
of the specific heat \cite{Idiaszek03}, to define a characteristic temperature which 
approaches the critical temperature for $N \rightarrow \infty$.
Numerically we find that even for small clusters
containing a few bosons 
$\rho_c^{\rm BEC} (N) \approx \rho_c^{\rm SSB} (N)$.
The most efficient way to obtain the Landau functions
is to directly calculate $P_N^{\rm BEC} ( q )$ 
recursively~\cite{Balazs98}, and then use
Eqs.~(\ref{eq:Lphires},\ref{eq:PNSSB},\ref{eq:PNBEC}).
As expected, with increasing $N$ 
the critical density  slowly approaches the bulk value
$\rho_c = \zeta ( 3 ) \approx 1.09$ from above
(for example
$\rho_c (100) \approx 1.93$ and $\rho_c (500) \approx 1.60$), and 
the probability distributions develop narrows peaks, so that the difference
between the most probable and the average value 
of the order parameter vanishes. 
The advantage of parameterizing the probability distributions
in terms of  Landau functions is that
only their detailed shape (but not their overall scale) depends on $N$, 
because the leading $N$-dependence has already been taken into account via 
the prefactor $N$ in the exponent $\exp [ - N {\cal{L}}_N ]$.

In the rest of this work we consider
the limit $ N \rightarrow \infty$ with fixed $\rho$.
For FB this is the usual thermodynamic limit, while for HB
this limiting procedure
amounts to letting  $\omega \rightarrow 0$ at
fixed $\omega^D N \propto \rho$, see Ref. [\onlinecite{Pitaevskii03}].
Then the summation in Eq.~(\ref{eq:Lphires}) is sharply
peaked at $ n =  N  | \phi |^2$, so that
$ {\cal{L}}^{\rm SSB}_\infty ( \phi )
=   {\cal{L}}^{\rm BEC}_\infty ( |\phi |^2 )
 \equiv {\cal{L}}_{\infty} ( | \phi |^2 )$ where
 $
 {\cal{L}}_{\infty} ( q  ) = - \lim_{ N \rightarrow \infty} N^{-1} \ln I_N ( q )
 $,
and $I_N ( q )$ is
the following integral in the complex $z = x + i y$ plane,
 \begin{equation}
 I_N ( q ) 
 = \int_{ 0 }^{2 \pi i } \frac{ d z }{2 \pi i }  e^{   N f ( z ) }
 \; .
 \label{eq:IN}
 \end{equation}
The complex function $f(z)$ is in $D$ dimensions given by 
 \begin{equation}
 f ( z ) = z ( 1-q ) +  \rho^{-1} g_{d +1 } ( z )    
\; .
 \label{eq:fdef}
 \end{equation}
Recall that
 $d = D/2$ for FB and $d=D$ for HB.
The Bose-Einstein integral $g_s (z)$  
is defined  by \cite{Robinson51} 
 \begin{equation}
 g_{s} ( z  ) 
 = \frac{1}{\Gamma ( s )}
 \int_0^{\infty} du   \frac{ u^{s-1} }{ e^{ u + z} - 1 }
 \; ,
 \label{eq:gsdef}
 \end{equation}
were $\Gamma (s)$ is the $\Gamma$-function.
In the complex $z$-plane  $g_s ( z )$
has infinitely many branch points at integer
multiples of $2 \pi i$, with branch cuts parallel to the negative real axis.

\begin{figure}[tb]    
   \centering
 \psfrag{y}{$y/ \pi$}
 \psfrag{x}{$x$}
 \psfrag{Ref}{$ {\rm Re} f $}
   \epsfig{file=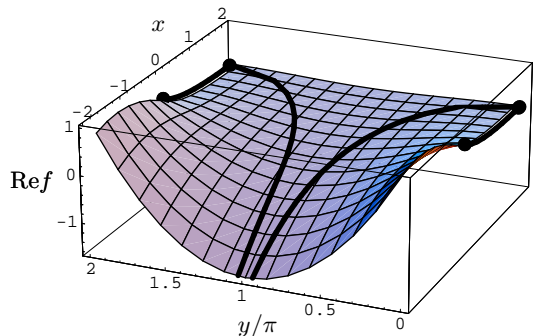,width=70mm}
  \vspace{-4mm}
  \caption{%
(Color online) Typical behavior of 
${\rm Re } f ( x+iy )$ 
defined in Eq.~(\ref{eq:fdef}) in the regime $q > q_c$.
The graph is for $d=3/2$, 
$q =0.95$, and $q_c =0.05$.
The end points of the integration and the saddle points
are marked by black  dots.
The thick line marks the value of ${\rm Re} f$ on  a deformation of the
original integration contour in Eq.~(\ref{eq:IN}) 
along steepest descend paths 
through the saddle points $z_0 = \alpha ( q , \rho)$ and
$z_1 = \alpha (q,\rho) + 2 \pi i$. 
}
    \label{fig:descend1}
  \end{figure}

The leading asymptotic behavior of the  $I_N ( q )$ for large
$N$ can be obtained using the 
method of steepest descends \cite{Bleistein86}.
For
 $\rho < \rho_c = g_{ d } (0) = \zeta ( d )$, or
if $ \rho > \rho_c$ but
$q > q_c = 1 - \rho_c / \rho > 0 $,
the saddle point equation
\begin{equation}
 d f ( z )/{ d z } = 1 - q -  \rho^{-1} g_{d} (z )  =0
\label{eq:saddle} 
\end{equation}
has a solution $z_0 = \alpha ( q, \rho)$ on the positive
real axis. 
Using the analyticity of the integrand, we may then deform the
original integration contour into a combination of
steepest descend paths
through $z_0$ and the
related saddle point $z_1 = \alpha ( q , \rho ) + 2 \pi i$,
as shown in Fig.~\ref{fig:descend1}.
The  asymptotic behavior of $I_N ( q )$
for $ N \rightarrow \infty$ is determined by 
the value of the integrand at the
saddle points $z_0$ and $z_1$, 
which yield complex conjugate contributions.
We thus obtain in this regime
 ${\cal{L}}_{\infty} ( q ) = - f ( \alpha ( q , \rho) ) $.
The corresponding effective potential 
$U_{\rm eff} ( \phi ) =   {\cal{L}}_{\infty} (  | \phi |^2 ) -   {\cal{L}}_{\infty} ( 0)$ 
shows rather non-trivial behavior,
in spite of the fact that we are dealing with 
a non-interacting system.
For simplicity, we consider here only the regime 
close to the phase transition, where
$| \rho - \rho_c | \ll  \rho_c$.
In the normal phase $\rho < \rho_c$
we may expand $U_{\rm eff}$ 
in powers of $| \phi |^2$,
\begin{equation}
 U_{\rm eff} ( \phi ) = \alpha_0  | \phi |^2 + 
\frac{ u_0}{2} | \phi |^4
 + O ( | \phi |^6 )
 \; .
 \label{eq:Ueffdef}
 \end{equation}
Here $\alpha_0  = \alpha ( q=0, \rho) $  
is the real solution of 
Eq.~(\ref{eq:saddle}) for $q =0$, i.e. 
 $
 \rho = g_{d} ( \alpha_0 )
 $, and $u_0  =  \chi^{-1} ( \alpha_0 )$, where
$\chi  ( \alpha ) = \rho^{-1} g_{ d -1 } ( \alpha )$
is the susceptibility of the Bose gas.

Eq.~(\ref{eq:Ueffdef}) is the usual  behavior of the
effective potential in the vicinity of  continuous phase transitions. 
For $ \rho \rightarrow \rho_c$
the parameter $\alpha_0$ vanishes, so that at the first sight it seems that the
leading term of the effective potential at the critical point
is proportional to $ | \phi |^4$.
However, this conventional scenario is only correct
for $d > 2$, i.e. for dimensions $D > 4$  (FB) or
$D > 2$ (HB). 
For $d \leq 2$  
the susceptibility $\chi ( \alpha )$ 
diverges for $\alpha \rightarrow 0$, implying
$u_0 \rightarrow 0$ for $\rho \rightarrow \rho_c$.
For $1 < d < 2$ we find at the critical point to leading order
 $
 U_{\rm eff} ( \phi ) \approx A ( \rho_c ) | \phi |^{ \frac{2d}{d-1}}
 $
with
$
 A ( \rho )  = 
(1 - d^{-1}) 
    |  \rho  /     \Gamma ( 1 - d) |^{ \frac{1}{d-1}} 
$.
In particular, for FB in $D=3$ we obtain
 $U_{\rm eff} ( \phi ) = \frac{ [ \zeta ( 3/2) ]^2}{12 \pi}  | \phi |^{6}$.
In the condensed phase $\rho > \rho_c$ the Landau function for $q > q_c$
is still determined by the above saddle points $z_0$ and $z_1$.
In this case we obtain close to the phase transition
 \begin{equation}
 U_{\rm eff} ( \phi ) 
 \approx \left\{ 
 \begin{array}{ll}
\frac{u_0 }{2} ( | \phi |^2 - q_c  )^2 & \mbox{for $  d > 2$}
 \\
 A ( \rho )  ( | \phi |^2 - q_c )^{ \frac{d}{d-1}} & \mbox{for $ 2 > d > 1$}
 \end{array} \right.
\; .
 \label{eq:Ueff2}
 \end{equation}
Note that for $q >q_c$ the Landau function
can be identified with the Gibbs potential for $N \rightarrow \infty$, 
so that the above results can also be derived
from thermodynamics. 

On the other hand,  in the condensed phase $\rho > \rho_c$
the Gibbs potential is constant for
$q < q_c = 1 - \rho_c/ \rho$, and is not necessarily equal to the 
Landau function~\cite{Goldenfeld92}.
It turns out that in this regime the saddle point
equation (\ref{eq:saddle}) does not have solutions on the
principal sheet of the Riemann surface of the complex function
$f(z)$ defined in Eq.~(\ref{eq:fdef}).
However, there are saddle points on higher sheets, two of which
(which we call $\tilde{z}_0$ and $\tilde{z}_1$)
are the analytic continuation of the
saddle points $z_0$ and $z_1$ discussed above for $ q >q_c$.
While we have not attempted a global analysis of the descend paths
on the higher sheets of the Riemann surface of $f ( z )$ for
arbitrary $d$, we can show that 
for $ d > 5/3$  the  path connecting the origin to 
$\tilde{z}_0$
continues to be a steepest descend path, so that
the saddle points $\tilde{z}_0$ and $\tilde{z}_1$ are still
relevant for the asymptotics  of $I_N ( q )$ 
for $N \rightarrow \infty$.
The leading behavior of ${\cal{L}}_{\infty} (q)$ for small $ q_c -q > 0$ 
can then be obtained  
by analytic continuation of the corresponding result for $ q >q_c$, which amounts to replacing
$  | \phi |^2 - q_c $ by $  q_c - | \phi |^2$ in Eq.~(\ref{eq:Ueff2}).
Hence,  for $ d > 5/3$ the effective potential 
qualitatively resembles the ``Mexican hat''
characteristic for continuous phase transitions, 
although for $5/3 < d < 2$ its shape
cannot be approximated by a quartic potential~\cite{Ledowski04}.

\begin{figure}[tb]    
   \centering
 \psfrag{Ref}{${\rm Re} f $}
 \psfrag{y}{$y / \pi$}
 \psfrag{x}{$x$}
      \epsfig{file=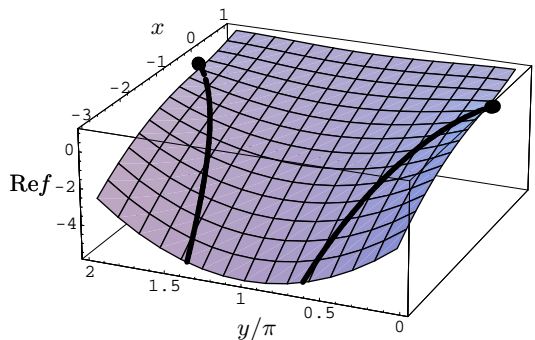,width=70mm}
  \vspace{-4mm}
  \caption{%
(Color online) ${\rm Re} f ( x+iy )$ and steepest descend paths
for $q < q_c$. The graph is for $d=3/2$, $q_c =0.05$, and $q =0.045$.
}
    \label{fig:descend2}
  \end{figure}

On the other hand, for $1 < d < 5/3$
the saddle points $\tilde{z}_0$ and $\tilde{z}_1$ become 
{\it{inadmissible}} for the asymptotic expansion
of $I_N ( q )$ in the sense defined on
p. 267 of Ref. [\onlinecite{Bleistein86}];  the end-points
of the integration can then be directly connected by steepest descend
paths avoiding any saddle point.  
For $d=3/2$ the proper deformation 
of the integration contour
is shown in Fig.~\ref{fig:descend2}.
As a consequence, for $d < 5/3$ the Landau function 
is constant ${\cal{L}}_{\infty} ( q )  = - \rho^{-1} g_{ d +1 } (0)$
for $0< q <  q_c = 1 - \rho_c / \rho$, just like the Gibbs potential.
Note that in 
$D=3$  the behavior of ${\cal{L}}_{\infty} (q )$
depends on the form of the external potential:
while for FB ($d = D/2 = 3/2$) the Landau function
has a plateau for $0< q <  q_c$,  for HB
($d = D =3$) it
has the ``Mexican hat'' form.

In summary, we have defined and evaluated the Landau functions and
the order parameter probability distributions for BEC and SSB
of non-interacting bosons in a canonical ensemble.
The shape of the Landau functions in the condensed phase depends
on the dimensionality and on the shape of
an external potential in a surprisingly complex way.
For bosons in a harmonic trap the
evolution of the Landau functions with density or temperature
is typical for second order phase transitions.
The shape  of 
Landau functions leads to the most natural extension of the
concept of the critical temperature for BEC and SSB in finite systems.


We thank N. Hasselmann and L. Banyai for discussions, and E. J. Mueller for
his comments and for
pointing out some relevant references.


%

\end{document}